\documentclass[fleqn,usenatbib]{mnras}

\usepackage{newtxtext,newtxmath}

\usepackage[T1]{fontenc}

\DeclareRobustCommand{\VAN}[3]{#2}
\let\VANthebibliography\thebibliography
\def\thebibliography{\DeclareRobustCommand{\VAN}[3]{##3}\VANthebibliography}

\usepackage{graphicx}	
\usepackage{amsmath}	
\usepackage{xcolor}

\defcitealias{Scowcroft2016_CO}{S16}

\title[CO in Cepheid Variables]{First detection of CO emission from Cepheid Variable stars.}

\author[S. L. Hamer et al.]{
S. L. Hamer,$^{1}$\thanks{E-mail: stephen.hamer.astro@gmail.com (Hamer)}
S. Ardern,$^{1}$
V. Scowcroft$^{1}$
\\
$^{1}$Department of Physics, University of Bath, Claverton Down, Bath, BA2 7AY, UK\\
}

\date{Accepted XXX. Received YYY; in original form ZZZ}

\pubyear{2015}

\begin{document}
\label{firstpage}
\pagerange{\pageref{firstpage}--\pageref{lastpage}}
\maketitle

\begin{abstract}
We present IRAM 30~m Telescope observations of the CO(1-0) and CO(2-1) emission lines in a sample of eight Cepheid variable stars.  The CO(1-0) line is detected in four of the eight targets at a signal-to-noise of $>$3.5 confirming the presence of CO in Cepheid atmospheres.  Two sources show strong absorption in both CO lines; this is likely related to contamination by cold molecular gas clouds along or close to the line of sight.  The remaining two targets showed no strong features related to either CO line.   These detections represent the first direct evidence for the presence of CO in Cepheid atmospheres, providing strong evidence for the mechanism proposed to explain the observed mid-infrared colour variation seen in Cepheids. Further, these detections support the proposed use of mid-IR colour as a robust photometric metallicity indicator for Cepheids, potentially leading to the elimination of metallicity systematics from the $H_0$ error budget. We discuss the future studies needed in this area and how our observations can be used to inform optimal observing strategies for large-scale dedicated studies.

\end{abstract}




\section{Introduction}
Modern Cosmology has found itself at an impasse.  The latest and most precise estimates of the Hubble constant ($H_0$) obtained from {\em Planck} (`Standard Ruler' measurements) \citep{Planck2020} disagree at the 4.2\,$\sigma$ level with those obtained from the cosmic distance ladder using `Standard Candles' \citep[e.g.][]{Freedman2012, Riess2022}.  This disagreement, dubbed the `Hubble Tension', is one of the greatest challenges facing modern cosmology.  A true divergence of these measurements would require us to reevaluate key models underpinning our understanding of the Universe, such as the $\Lambda$CDM model and the number of neutrinos in the standard model \citep{Bernal2016, DiValentino2017}.  Given these implications, it is essential to rule out systematic uncertainties in either experiment as the source of the Hubble tension.

The majority of distance ladder studies use Cepheids in the Milky Way and/or Magellanic Clouds as their `bottom rung'. In recent years, the shift to near- and mid-infrared wavelengths, where the intrinsic dispersion in the Leavitt law (LL) is smaller, has improved the error budget for Cepheid distances \citep[e.g.][]{Freedman2012, Riess2022}, though the impact of metallicity \citep[][and references therein]{Ripepi2020, Romaniello2022} and the potential non-linearity of the LL remain open questions \citep[e.g.][]{Kodric2015,Chown2021}. These must be answered before the question of the Hubble tension can be addressed, as systematics at the base of the distance ladder will be compounded with each further step towards $H_0$ \citep{Freedman2021}.

In a \textit{Spitzer}-IRAC study of Galactic and Magellanic Cloud Cepheids, \citet[][hereafter S16]{Scowcroft2016_CO} found that Cepheids with $0.8 \lesssim \log P \lesssim 1.8$~d exhibit a clear period-colour (PC) relation in the mid-infrared (mid--IR), with longer period objects having bluer colours. \citetalias{Scowcroft2016_CO} hypothesised that the cyclical variation in mid-IR colour (their figure 2) and the PC relation are due to the significant CO vibration-rotation band-head at $4.6~\mu$m, first identified in stars by \citet{Ayres1994} and \citet{Wiedemann1994}. At the typical temperature range of Cepheids ($\sim$ 4000-6000\,K) CO is readily destroyed by chemical processes such as H + CO $\rightarrow$ C + OH.  The rate of these reactions varies strongly within the temperature range of Cepheids, from $10^{-17}$ to $10^{-15}~\text{cm}^{3}~\text{s}^{-1}$ at their coolest and hottest temperatures, respectively, while the formation rate of CO through radiative association remains relatively constant at 10$^{-16}$\,cm$^{3}$\,s$^{-1}$.

The effect of metallicity on Cepheid luminosities is expected to be reduced at longer wavelengths. However, \citetalias{Scowcroft2016_CO} identified a significant metallicity dependence in their mid-IR colours. 
Combining their \textit{Spitzer} photometry with spectroscopic metallicities from \citep{Genovali2014}, \citetalias{Scowcroft2016_CO} showed that the deviation from the LMC PC can be used as a photometric metallicity indicator, producing [Fe/H] estimates with accuracies competitive with spectroscopic measurements \citepalias[their figure 7]{Scowcroft2016_CO}. As one of the challenges faced by previous metallicity studies has been the limited number of Cepheids for which accurate stellar metallicity measurements are available, a robust photometric metallicity indicator for Cepheids would be a significant step forward in elimitating the metallicity systematic from the distance ladder. 

Here we present the first results of an IRAM-30m pilot study to confirm the validity of the CO mechanism that \citetalias{Scowcroft2016_CO} propose as the source of the mid-IR colour-metallicity relation. We show for the first time that CO is present in Cepheid atmospheres, providing observational evidence for the cyclical CO destruction and formation mechanism discussed in \citetalias{Scowcroft2016_CO}, further strengthening the case for the use of mid-IR colour as a metallicity indicator for Cepheids. 


\section{Sample}

Prior to the observing run, a subset of 52 Galactic Cepheids with \textit{Gaia} time-series photometry \citep{Gaia_2016, Gaia_2022_DR3} was identified, with the condition that each target would be observable at least 50\% of the time during the IRAM run, and that mid-IR photometry from \citet{Monson2012} and/or spectroscopic metallicities from \citet{Genovali2014} were available.

On each day of observations, targets were selected and prioritised based on the predicted phase from the \textit{Gaia} light curve at the time of observation.  Targets approaching minimum light were selected; according to \citetalias{Scowcroft2016_CO}'s proposed mechanism the highest CO production is expected in this region, thus these phases represented the highest potential for CO detection.  A secondary cut, removing sources close to the galactic plane was implemented to avoid contamination of the beam by intervening gas clouds. Finally, each candidate source was required to have a bright pointing source within 20 arcsec at the time of the observations to ensure accurate pointing of the telescope and maximise the sensitivity to emission from the source.

\section{Observations}

The observations were carried out with the IRAM 30m telescope at Pico Veleta, Spain, in one run during 8$^{\text{th}}$ -- 12$^{\text{th}}$ February 2023. Observations of the CO(1-0) and CO(2-1) lines were conducted simultaneously using the E090 and E230 receivers respectively in parallel.  The broadband EMIR receivers were used to observe in dual polarisation mode using a single side band and a total bandwidth of 4 GHz.  Wobbler switching mode was used for background subtraction with a phase of 2 seconds.  
Due to technical issues with the Wobbler the throw had to be reduced from 60 to 30 arcsecs for the final two days of observations.
The receivers were tuned to the rest frequency of CO(1-0) \& CO(2-1) (115.27 and 230.54 GHz respectively) and the tuning was tested each day by observing the line calibration source W3(OH).  The detected fluxes and peak antenna temperature measured for this source are consistent with the values presented in the line calibration catalogue \citep{mbg98} and were consistent between observing days to within the accepted tolerance of the telescope ($<$10\% variation, Table \ref{tab:line_cal}).

\begin{table}
    \caption{Peak and total line flux measured for the line calibrator (W3HO) on each day of observations.  Reference values peak at $\sim$23\,K (181.7\,Jy) and $\sim$18\,K (180.0\,Jy) respectively \citep{mbg98}.  Our measured peaks are consistent with these values on each day and the total flux in the lines varies by $\sigma$=4.5\% (max=10.3\%) and $\sigma$=2.9\% (max=6.2\%) of the mean.}
    \label{tab:line_cal}
    \centering
    \begin{tabular}{c|c|c|c|c}
    \hline
    {\bf Date} & {\bf CO(1-0)} & {\bf CO(1-0)} & {\bf CO(2-1)} & {\bf CO(2-1)} \\
    & {\bf Peak} & {\bf Flux} & {\bf Peak} & {\bf Flux} \\
    \hline
     08/02 & 26.8\,K & 258\,K\,km\,s$^{-1}$ & 21.5\,K & 297\,K\,km\,s$^{-1}$ \\
     & 161.3\,Jy & 1553\,Jy\,km\,s$^{-1}$ & 129.4\,Jy & 1788\,Jy\,km\,s$^{-1}$ \\
     09/02 & 27.6\,K & 263\,K\,km\,s$^{-1}$ & 23.9\,K & 306\,K\,km\,s$^{-1}$ \\
     & 166.2\,Jy & 1583\,Jy\,km\,s$^{-1}$ & 143.9\,Jy & 1842\,Jy\,km\,s$^{-1}$ \\
     11/02 & 26.2\,K & 251\,K\,km\,s$^{-1}$ & 22.6\,K & 316\,K\,km\,s$^{-1}$ \\
     & 157.7\,Jy & 1511\,Jy\,km\,s$^{-1}$ & 136.0\,Jy & 1902\,Jy\,km\,s$^{-1}$ \\
     12/02 & 25.3\,K & 237\,K\,km\,s$^{-1}$ & 21.3\,K & 298\,K\,km\,s$^{-1}$ \\
     & 152.3\,Jy & 1426\,Jy\,km\,s$^{-1}$ & 128.2\,Jy & 1794\,Jy\,km\,s$^{-1}$ \\
    \hline    
    \end{tabular}
\end{table}

The FTS200 and WILMA backends were used in parallel to sample the data at 0.2 and 2\,MHz respectively.  Due to technical issues, poor weather, and the availability of pointing sources total integration times per source differed significantly.  As we expect significant variability of each source over its pulsation period, stacking observations from different days was not considered a viable solution to increase observation depth. We present the total integration time per source, the RMS noise reached, and the average atmospheric opacity (Tau) during each source's observations in Table. \ref{tab:sou}.  Pointing \& Focus measurements were repeated at least every 1.5 hours. 

The line emission is measured in antenna temperature T$^*_A$ and converted to fluxes (Jy) using the point source sensitivity conversion S/T$_A^*$ = 6.02\,Jy/K (half power beam width of 21.4 arcsec at 115\,GHz) and S/T$_A^*$ = 7.8\,Jy/K (half power beam width of 10.7 arcsec at 230\,GHz) for the E090 and E230 bands respectively. The data were reduced with the \textsc{class/gildas} software, and the spectra were smoothed to a maximum of $\sim$40km\,s$^{-1}$ consistent with the maximum linewidth expected to result from the movement of the stellar surface as it pulsates. We expect our sources to have small line of sight velocities so set our velocity zero-point at the rest frame frequency of the respective lines. 

\begin{table*}
    \caption{Details of the observations. T$_{\text{int}}$ is the total integration time spent observing the source. $\Delta$V is the velocity resolution used for smoothing, with a maximum of $\sim$40km\,s$^{-1}$ used to be consistent with expected linewidths induced by Cepheid pulsations. 
    RMS is the noise level obtained at the given smoothing. Opacity is the mean Tau value measured during the observations.  All observation dates are 2023.}
    \label{tab:sou}
    \centering
    \begin{tabular}{c|c|c|c|c|c|c|c|c|c|c|c}
    \hline
    {\bf Source} & {\bf T$_{\text{int}}$} & {\bf Backend} & $\Delta$V & \multicolumn{2}{c}{\bf RMS \@ 115\,GHz} & \multicolumn{2}{c}{\bf RMS \@ 230\,GHz} & \multicolumn{2}{c}{\bf Opacity (Tau)} & {\bf Date} & {\bf Period}\\
    & {\bf (min)} & & {\bf (km\,s$^{-1}$}) & {\bf (mK)} & {\bf (mJy)} & {\bf (mK)} & {\bf (Jy)} & {\bf 115\,GHz} & {\bf 230\,GHz} & & {\bf Days} \\
    \hline
     GY~Sge & 40.2 & FTS & 32.5 & 2.93 & 17.6 & 3.04 & 23.9 & 0.373 & 0.232 & 08/02 & 51.05 \\
     & & WILMA & 41.6 & 2.28 & 13.8 & 3.07 & 23.9 & & & \\
     BZ~Cyg & 78.5 & FTS & 32.5 & 1.58 & 9.5 & 1.16 & 9.0 & 0.378 & 0.242 & 08/02 & 10.14 \\
     & & WILMA & 41.6 & 1.13 & 6.8 & 1.07 & 8.3 & & & \\
     CH~Cas (1) & 40.3 & FTS & 32.4 & 2.12 & 12.8 & 1.95 & 15.2 & 0.380 & 0.252 & 08/02 & 15.09 \\
     & & WILMA & 41.6 & 1.69 & 10.2 & 1.52 & 11.9 & & & \\
     CP~Cep & 91.4 & FTS & 32.5 & 1.67 & 10.1 & 0.93 & 7.3 & 0.345 & 0.072 & 11/02 & 17.86 \\
     & & WILMA & 41.6 & 1.33 & 8.0 & 0.67 & 5.3 & &  \\
     RY~Cas & 81.1 & FTS & 32.5 & 1.64 & 9.9 & 0.83 & 6.5 & 0.344 & 0.079 & 11/02 & 12.14 \\
     & & WILMA & 41.6 & 1.29 & 7.8 & 0.89 & 6.9 & &  \\
     KX~Cyg & 50.7 & FTS & 32.5 & 1.79 & 10.8 & 1.01 & 7.9 & 0.343 & 0.075 & 11/02 & 20.04 \\
     & & WILMA & 41.6 & 1.75 & 10.5 & 1.53 & 11.9 & &  \\
     CH~Cas (2) & 61.0 & FTS & 32.5 & 2.04 & 12.3 & 2.30 & 18.5 & 0.420 & 0.363 & 12/02 & 15.09 \\
     & & WILMA & 41.6 & 1.98 & 11.9 & 2.00 & 15.6 & &  \\
     CY~Cas & 61.0 & FTS & 16.3 & 3.21 & 19.4 & 3.52 & 27.5 & 0.403 & 0.327 & 12/02 & 14.38 \\
     & & WILMA & 20.8 & 3.04 & 18.3 & 2.71 & 21.1 & &  \\
     V396~Cyg & 91.3 & FTS & 32.5 & 1.07 & 6.4 & 1.48 & 11.5 & 0.398 & 0.325 & 12/02 & 33.25 \\
     & & WILMA & 41.6 & 1.07 & 6.4 & 1.43 & 11.2 & &  \\

    \hline    
    \end{tabular}
\end{table*}

\section{Results}
\subsection{CO detections}

\begin{figure*}
    \centering
    \includegraphics[width=0.995\textwidth]{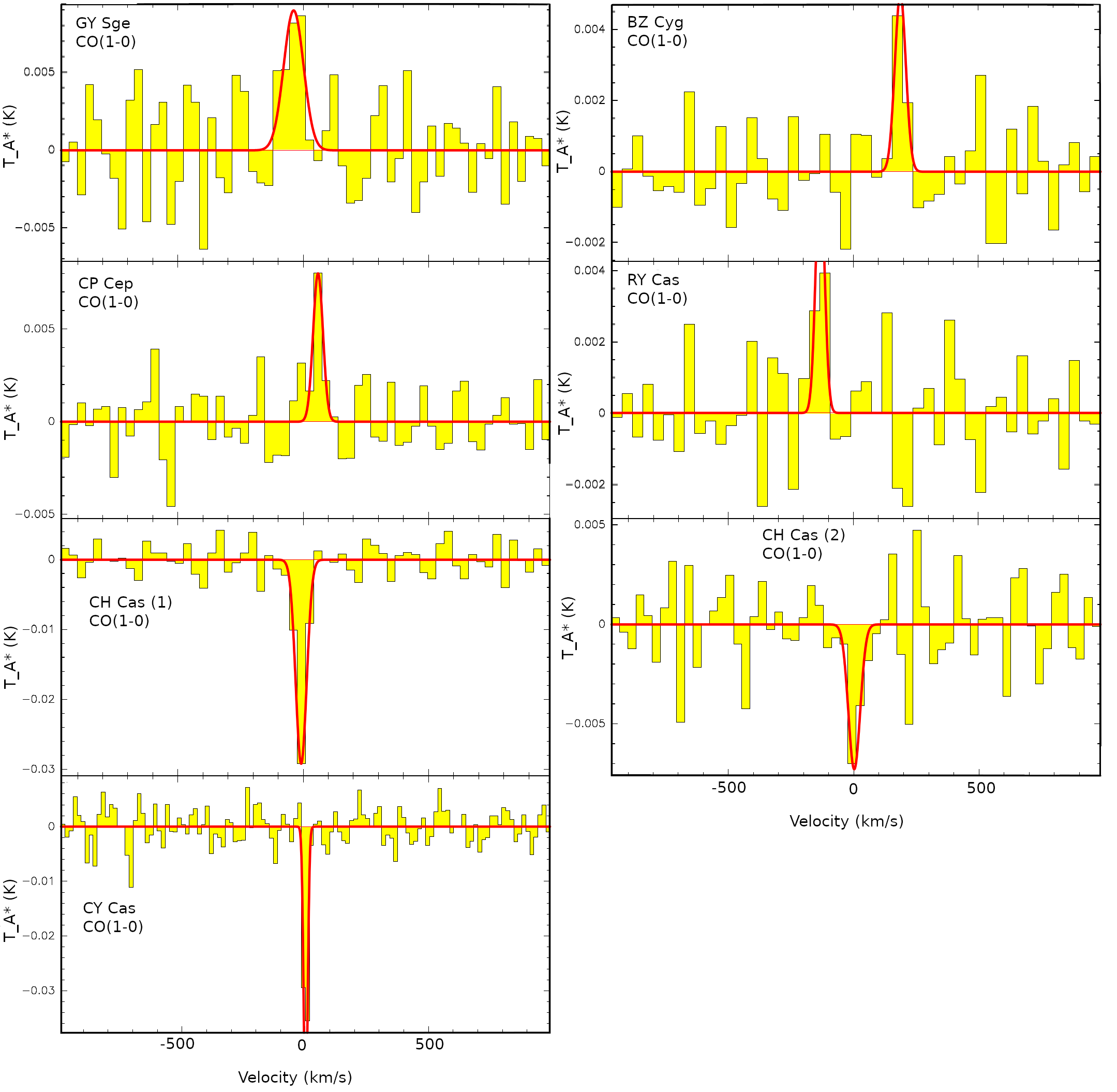}
    \vspace{-4.0cm}
    
    \begin{minipage}[b]{0.55\textwidth}
      \end{minipage}
      \hfill
    \begin{minipage}[b]{0.45\textwidth}
    \caption{Plots showing all of our CO(1-0) detections at a signal-to-noise of greater than 3 produced using \textsc{class/gildas} for the backend which provided the highest signal-to-noise.  The individual plots are labelled to indicate which observation/object each refers to.  The yellow bars show she observed spectra in antenna Temperature (K), smoothed to the velocity resolution indicated in Table \ref{tab:sou}.  Fits to detected features are shown as a red line, the parameters of these fits can be found in Table \ref{tab:detec}. The repeat observations of CH~Cas is shown next to the first observation despite not reaching a signal-to-noise of $>$3 for comparison. }
    \label{fig:detections}
    \end{minipage}
    \vspace{0.4cm}
    
\end{figure*}

Of the eight Cepheid Variables observed, four were detected in emission at $>$3$\sigma$ significance in at least one of the backends (see Table \ref{tab:detec}).  A further two showed unexpected strong absorption features, the nature of which is discussed in detail in \S \ref{sec:abs}.  Just one source (CH~Cas) had repeat observations on multiple days to check for variability, this is discussed in \S \ref{sec:var}.  Spectra for each source detected are presented in Figure \ref{fig:detections} showing the CO(1-0) emission as seen in the backend that provided the highest signal-to-noise detection.  

Considering the four sources detected in emission, they show an average flux of 2.9\,Jy\,km\,s$^{-1}$ with a range of 1.78 -- 5.37\,Jy\,km\,s$^{-1}$.  Of these GY~Sge has by far the highest flux at more than twice that measured for the other three sources.  In line with the model presented in \citetalias{Scowcroft2016_CO} CO production is expected to start at a phase of $\sim$0.1 and continue until $\sim$0.65 after which it is quickly destroyed as the light curve rises again.  Because of this cyclical formation and destruction of CO Cepheids with longer periods will have substantially more time to build up CO in their atmospheres than Cepheids with shorter periods.  As such GY~Sge having significantly more flux than the other detected sources is consistent with the model presented by \citetalias{Scowcroft2016_CO} as its period is significantly longer (51.05 days) than any of the other detected sources (CP~Cep is the next longest with 17.86 days).  

GY~Sge is also observed at a phase of 0.25 putting it on the falling part of its light curve where CO production is expected to be at a maximum in line with the model of \citetalias{Scowcroft2016_CO}.  By contrast, BZ~Cyg, CP~Cep, and RY~Cas are all observed close to the bottom of their optical light curves where CO production is expected to transition to CO destruction.  As there remains some ambiguity in the exact point at which this transition occurs (it is related to the star's temperature variations rather than its flux variations so it is not possible to make direct comparisons between stars with different periods) it is possible that some of these sources where caught during the rapid CO destruction phase resulting in lower measured fluxes.   In order to fully characterise this cycle of CO creation and destruction it will be necessary to regularly monitor the CO flux from Cepheid Variables with a range of periods throughout the entirety of their period, and ideally over multiple pulsations.

All detections have velocity measurements that are close to rest (0\,km\,s$^{-1}$) and are well within the typical range found for Cepheids within the {\em Gaia DR3} catalogue (100--300\,km\,s$^{-1}$).  The measured FWHM for BZ~Cyg, CP~Cep, and RY~Cas are consistent with the maximum line with expected to be induced by stellar pulsations (FWHM$\sim$40\,km\,s$^{-1}$).  GY~Sge has a FWHM of 93\,km\,s$^{-1}$ which is considerably higher than this. It is also substantially higher than the majority of measured line broadening parameters for Cepheids in the {\em Gaia DR3} catalogue (typically $<$50\,km\,s$^{-1}$) though it remains well within the range measured ($<$200\,km\,s$^{-1}$).  The subtracted baseline was checked to ensure that an undersubtraction in the region of the line was not responsible for the detected feature, but the baseline was flat and constant over the range (200--100\,km\,s$^{-1}$) indicating that the feature must have been present prior to the baseline subtraction.

As such the four sources GY~Sge, BZ~Cyg, CP~Cep, and RY~Cas all show significant detections of CO(1-0) making them the first detections of CO emission from Cepheid Variable Stars.  RY~Cas also shows CO(2-1) emission detected at 3.8$\sigma$ with a velocity and FWHM consistent with the CO(1-0) line.  The none detections of KX~Cyg and V396~Cyg can potentially be attributed to the short integration time of KX~Cyg and the extremely poor conditions in which V396~Cyg was observed.  Regardless, these detections represent an important result in the field indicating that CO is indeed present in the atmospheres of Cepheid Variables and can thus explain the colour variations seen by \citetalias{Scowcroft2016_CO}.

\begin{table*}
    \caption{Details of the detections made for each source.  Details of both backends are listed for comparison when a $>$3$\sigma$ detection was made with either.  The parameters of Gaussian fits to the line features are listed.  Some lines where unresolved at the binning used, as indicated by an error on the FWMH which is significantly greater than the measured FWHM. Phase is the point in the light curve at which the observation was taken, measured from the peak.}
    \label{tab:detec}
    \centering
    \begin{tabular}{c|c|c|c|c|c|c|c|c|c|c}
    \hline
    {\bf Source} & {\bf Line} & {\bf Backend} & \multicolumn{2}{c}{\bf Flux} & {\bf Velocity} & {\bf FWHM} & \multicolumn{2}{c}{\bf Peak} & {\bf Phase} & {\bf S/N} \\
    & & & {\bf (K\,km\,s$^{-1}$)} & {\bf (Jy\,km\,s$^{-1}$)} & {\bf (km\,s$^{-1}$)} & {\bf (km\,s$^{-1}$)} & {\bf (mK)} & {\bf (mJy)} &  {\bf $\phi$} & \\ 
    \hline
    GY~Sge & CO(1-0) & FTS & 0.89$\pm$0.23 & 5.37$\pm$1.39 & -38$\pm$13 & 93$\pm$25 & 9.0$\pm$2.9 & 54.1$\pm$17.6 & 0.25 & 3.87 \\
            &         & WILMA & 0.81$\pm$0.21 & 4.89$\pm$1.26 & -55$\pm$12 & 93$\pm$25 & 8.1$\pm$2.3 & 49.0$\pm$13.8 & & 3.86 \\
    BZ~Cyg & CO(1-0) & FTS & 0.35$\pm$0.13 & 2.11$\pm$0.78 & 190$\pm$15 & 80$\pm$35 & 4.1$\pm$1.58 & 24.7$\pm$9.5 & 0.48 & 2.69 \\
            &         & WILMA & 0.28$\pm$0.08 & 1.69$\pm$0.48 & 188$\pm$12 & 52$\pm$30 & 5.0$\pm$1.13 & 30.1$\pm$6.8 & & 3.50 \\
    CH~Cas (1) & CO(1-0) & FTS & -1.60$\pm$0.13 & -9.63$\pm$0.78 & -8$\pm$2 & 51$\pm$4 & -29.2$\pm$2.12 & -175.8$\pm$12.8 & 0.43 & 12.31 \\
            &         & WILMA & -1.58$\pm$0.14 & -9.51$\pm$0.84 & -8$\pm$2 & 62$\pm$10 & -23.7$\pm$1.69 & -142.7$\pm$10.2 & & 11.29 \\
            & CO(2-1) & FTS & -0.49$\pm$0.11 & -3.82$\pm$0.86 & -10$\pm$6 & 40$\pm$9 & -11.5$\pm$1.95 & -89.7$\pm$15.2 & & 4.45 \\
            &         & WILMA & -0.38$\pm$0.09 & -2.96$\pm$0.70 & -10$\pm$4 & 41$\pm$111 & -8.5$\pm$1.52 & -66.3$\pm$11.9 & & 4.22  \\
    CP~Cep & CO(1-0) & FTS & 0.39$\pm$0.103 & 2.35$\pm$0.62 & 59$\pm$7 & 46$\pm$12 &  8.00$\pm$1.67 & 48.2$\pm$10.1 & 0.70 & 3.79 \\
            &         & WILMA & 0.225$\pm$0.102 & 1.35$\pm$0.61 & 53$\pm$16 & 56$\pm$23 & 3.80$\pm$1.33 & 22.9$\pm$8.0 & & 2.21\\
    RY~Cas & CO(1-0) & FTS & 0.295$\pm$0.098 & 1.78$\pm$0.59 & -136$\pm$10 & 52$\pm$16 & 5.38$\pm$1.64 & 32.4$\pm$9.9 & 0.56 & 3.01\\
            &         & WILMA & 0.300$\pm$0.082 & 1.81$\pm$0.49 & -133$\pm$5 & 42$\pm$272 & 6.77$\pm$1.29 & 40.8$\pm$7.8 & & 3.66 \\
            & CO(2-1) & FTS & 0.208$\pm$0.055 & 1.62$\pm$0.43 & -143$\pm$8 & 58$\pm$16 & 3.38$\pm$0.83 & 26.4$\pm$6.5 & & 3.78 \\
            &         & WILMA & 0.147$\pm$0.057 & 1.45$\pm$0.44 & -138$\pm$6 & 42$\pm$239 & 3.32$\pm$0.89 & 25.9$\pm$6.9 & & 2.94 \\
    KX~Cyg & \multicolumn{8}{c}{No Detection} & 0.61 \\
    CH~Cas (2) & CO(1-0) & FTS & -0.43$\pm$0.15 & -2.59$\pm$0.90 & 2$\pm$8 & 55$\pm$27 & -7.28$\pm$2.04 & -43.8$\pm$12.3 & 0.75 & 2.87\\
            &         & WILMA & -0.44$\pm$0.17 & -3.43$\pm$1.32 & 19$\pm$17 & 75$\pm$31 & -4.85$\pm$1.98 & -37.8$\pm$18.5 & & 2.59 \\
            & CO(2-1) & & \multicolumn{4}{c}{\hspace{0.65cm} No Detection} \\
    CY~Cas & CO(1-0) & FTS & -1.09$\pm$0.08 & -6.56$\pm$0.48 & 2$\pm$1 & 19$\pm$9 & -53.88$\pm$3.21 & -324.4$\pm$12.3 & 0.29 & 13.63 \\
            &         & WILMA & -0.976$\pm$0.10 & -5.88$\pm$0.60 & 2$\pm$2 & 26 $\pm$10 & -35.68$\pm$3.04 & -214.8$\pm$11.9 & & 9.60 \\
            & CO(2-1) & FTS & -0.37$\pm$0.07 & -2.89$\pm$0.55 & 1$\pm$1 & 18$\pm$72 & -21.54$\pm$3.52 & -168.0$\pm$27.5 & & 5.29 \\
            &         & WILMA & -0.26$\pm$0.08 & -2.03$\pm$0.62 & 2$\pm$5 & 21$\pm$10 & -12.75$\pm$2.71 & -99.5$\pm$21.1 & & 3.25 \\
     V396~Cyg & \multicolumn{8}{c}{No Detection} & 0.74 \\
    \hline
    \end{tabular} 
\end{table*}

\subsection{Absorption features}
\label{sec:abs}
Of the eight sources observed two (CH~Cas and CY~Cas) showed exceptionally strong absorption features in both CO(1-0) and CO(2-1). 
Stars are not expected to have significant continuum emission at 115\,GHz and 230\,GHz \citep{pal96p} making it difficult to explain these features in the context of CO in the stellar atmospheres.  Likewise, an intervening gas cloud along the line of sight would still require a 'backlight' from the star to produce the absorption feature so suffers the same difficulty. While the spectra do show baselines that are sufficiently high to accommodate the absorption features the observations were not set up to detect continuum emission and there are many factors which can contribute to the baseline level so it is not possible to equate the baseline directly to continuum emission from the source.  

While the detections in both CO(1-0) and CO(2-1) make it unlikely, the individual scans were inspected one by one to rule out the possibility of a strong spike in one scan producing this feature, however evidence of the absorption feature was seen in each scan ruling out this possibility. An oversubtraction of the baseline was also ruled out as these strong features are clearly visible even prior to the baseline subtraction.  As such while this does not explain the origin of these absorption features, it confirms that they are real and not an artefact of the observations.

Considering the location of CH~Cas and CY~Cas, they are very close to each other on the sky and are both within the galactic plane of the galaxy.  This raises the possibility that the 'off' position used by the wobbler to remove the sky emission contained a Galactic molecular cloud with emission in CO(1-0) and CO(2-1) which was then subtracted from the emission from the source.   The features in both CH~Cas and CY~Cas have a velocity consistent with zero and show very narrow lines which is consistent with this hypothesis.  Additionally, the ratio of the CO(1-0) and CO(2-1) flux measurements is consistent with a higher population at J=1 than J=2, suggesting that the gas producing this feature is cold, and certainly colder than the 4000--6000\,K expected in the atmospheres of Cepheid Variables which rules out the absorption features coming from the stellar atmospheres.  As such a cold molecular cloud present at the wobblers 'off' position is the most likely explanation for these absorption features, though it is not possible currently to rule out an absorbing cloud along the line of sight.   \citet{jos98} encountered a similar issue and devised a method using position-switching mode to remove the absorption features which can be applied to future observations of Cepheids which show significant absorption.

\subsection{Variability}
\label{sec:var}

While ideally each Cepheid would have been observed at multiple epochs to test for variability during its pulsation cycle, due to poor weather and technical issues this was only possible for one source (CH~Cas).  Unfortunately, CH~Cas is one of the two sources which show CO in absorption which is difficult to attribute to CO in its stellar atmosphere.  Despite this the second observation of CH~Cas, taken four days after the first when the CO is expected to be in its destruction phase, shows significantly weaker features than on the first observation when it was well within the CO creation phase.  The CO(1-0) shows a day to day flux ratio of S$_{\phi = 0.43}$/S$_{\phi = 0.75}$ = 3.7 and the CO(2-1) is not detected at $\phi$ = 0.75.  While this is consistent with the expected behaviour of CO based on the model proposed by \citetalias{Scowcroft2016_CO} care must be taken when considering this variation as evidence that the CO absorption originates form the stellar atmosphere.  The two observations were taken at different times of day meaning the offset direction of the wobbler will have been different.  Due to technical issues with the wobbler, the throw had to be reduced from 60 to 30 arcsec between the two observations.  As a result of these two points in the scenario where the absorption features results from sky subtraction (see \S \ref{sec:abs}) the location of the offset position will be different for the two observations,  and the sky position may have caught a different region of the molecular gas cloud, resulting in different levels of 'absorption' being seen.

\section{Conclusions}

CO(1-0) emission is detected by IRAM 30m in four of the eight Cepheid variable stars targeted by this study representing the first direct detections of CO gas in Cepheid Variables. This provides strong evidence for the mechanism proposed by \citetalias{Scowcroft2016_CO} to explain the unexpected mid-IR colour curves, and further supports the use of mid-IR colours for direct metallicity estimates of Cepheids. Given the temperature of our sources, we expect the $4.6~\mu$m band-head absorption to be substantially stronger than our CO(1-0) detections and should be easily detected with {\em JWST} NIRSpec.
Two sources are detected in CO absorption, most likely caused by a molecular gas cloud being present at the offset location used for the sky subtraction.  To avoid this issue in future attempts to observe CO in Cepheids in the galactic plane we will use IRAM 30m's position switching mode, directed to a region of sky free from molecular gas clouds.  The detections of CO presented in this paper have important consequences for future projects in this field:  The measured fluxes will allow future studies at 1~mm \& 3~mm to more accurately estimate the observing time needed to detect emission from these stars. Finally, the confirmation of CO in Cepheid atmospheres highlights the need to study the variability of the absorption feature at 4.6$~\mu$m proposed by \citetalias{Scowcroft2016_CO} which can only be achieved with dedicated observations by {\em JWST}.

\section*{Data Availability}

The data underlying this article were obtained through observations at the IRAM 30m telescope (project 154-22).  Raw data are available from the IRAM Archive on direct request to IRAM, and metadata can be viewed through the TAPAS archive at \url{https://tapas.iram.es/tapas/}. Reduced data and spectra will be shared on reasonable request to the corresponding author.

\section*{Acknowledgements}
The team would like to thank the IRAM 30m Astronomer on Duty, Pablo Torne, for the extensive support and assistance they provided during the difficult conditions encountered during the observing run.  This publication has received funding from the European Union’s Horizon 2020 research and innovation programme under grant agreement No 101004719 (ORP). This work was supported by the Science and Technology Facilities Council [ST/S000526/1]. Ardern is supported by an STFC studentship.  



\bibliographystyle{mnras}
\bibliography{bib} 




\appendix




\bsp	
\label{lastpage}
\end{document}